\documentclass[twocolumn,showpacs,aps,prb]{revtex4-1}

\usepackage{amssymb,amsmath}
\usepackage[dvips]{graphicx}
\begin{document}

\title{Quantum Griffiths singularities in ferromagnetic metals}

\author{David Nozadze}
\author{Thomas Vojta}
\affiliation{Department of Physics, Missouri University of Science and Technology, Rolla, Missouri 65409, USA}

\begin{abstract}
We present a theory of the quantum Griffiths phases associated with the ferromagnetic quantum phase transition in disordered
metals. For Ising spin symmetry, we study the dynamics of a single rare region within the variational instanton approach. For Heisenberg
symmetry, the dynamics of the rare region  is studied using a renormalization group approach. In both cases,
the rare region dynamics is  even slower than in the usual quantum Griffiths case because the order parameter conservation of an
itinerant ferromagnet hampers the relaxation of large magnetic clusters. The resulting quantum Griffiths singularities in ferromagnetic
metals are stronger than power laws. For example, the low-energy  density of states $\rho(\epsilon)$ takes the asymptotic form $\exp[\{-\tilde{\lambda}\log (\epsilon_0/\epsilon)\}^{3/5}]/\epsilon$
with $\tilde{\lambda}$ being non-universal.
We contrast these results with  the antiferromagnetic case  in which the systems show power-law quantum Griffiths singularities
in the vicinity of the quantum critical point.  We also compare our result with existing
experimental data of ferromagnetic alloy ${\rm{Ni}}_{x}{\rm{V}}_{1-x}$.
\end{abstract}

\date{\today}

\maketitle

\section{Introduction}
The low-temperature behavior of  quantum many-particle systems can be sensitive to
impurities, defects, or other kinds of quenched disorder.  This effect
is especially important near  quantum phase transitions, where fluctuations
in time and space become connected. The interplay between  static disorder fluctuations
and large-scale quantum fluctuations  leads to much more dramatic effects at quantum phase
transitions than at classical phase transitions, including quantum Griffiths singularities, \cite{1969_Griffiths_PRL, 1995_Thill_Physyca_A, 1996_Rieger}
infinite-randomness critical points featuring exponential
rather than power-law scaling, \cite{1992_Fisher_PRL, 1995_Fisher_PRB} and the smearing of the transition.\cite{2003_Vojta_PRL}

The  Griffiths effects at a magnetic phase transition in a disordered system are caused by large spatial regions (rare regions)
that are devoid of impurities and can show local magnetic order even if the bulk system is
globally in the paramagnetic phase. The order parameter fluctuations induced by rare regions belong to
 a class of excitations known as {\it{instantons}}.  Their dynamics is very slow
because flipping the rare region  requires a coherent change of the order parameter over a large volume.
Griffiths  showed \cite{1969_Griffiths_PRL} that this leads to a singular free energy, not  just at  the transition point but
in a whole parameter region, which is now known as the Griffiths phase. In  classical systems,
the contribution of the rare regions to thermodynamic observables is very weak. However, due to the
perfect disorder correlations in (imaginary) time, Griffiths effects at quantum phase
transitions are  enhanced and lead to power-law singularities in thermodynamic quantities (for reviews see, e.g., Refs. \onlinecite{2006_Vojta_JPhysA, 2010_Vojta_JLTPhys}).

The systems in which quantum Griffiths behavior was originally demonstrated \cite{1995_Thill_Physyca_A, 1996_Rieger, 1992_Fisher_PRL, 1995_Fisher_PRB}
all have undamped dynamics (a dynamical exponent $z=1$ in the clean system). However,
many systems of experimental importance involve superconducting \cite{2003_Rogachev_APL} or magnetic
\cite{1994_Lohneysen_PRL, 2001_Grigera_Science, 1997_Pfleiderer_PRB,
1998_Andrade_PRL} degrees of
freedom coupled to conduction electrons. This leads to overdamped dynamics
characterized by a clean dynamical exponent $z>1$. Studying the effects of the rare regions
in this case is, therefore, an important issue. It has been shown that metallic Ising antiferromagnets can
show quantum Griffiths behavior at higher energies, where the damping is less important. \cite{2000_Castro_PRB}   In contrast, the
quantum Griffiths  singularities in Heisenberg antiferromagnets are caused by the dissipation and occur at lower energies. \cite{2005_Vojta_PRB}

In recent years, indications of  quantum Griffiths phases have been observed in experiments on a number of metallic
systems suchs as magnetic semiconductors,\cite{2007_Guo_PRL, 2010_Guo_PRB1, 2010_Guo_PRB2} Kondo lattice ferromagnets,  \cite{2007_Sereni_PRB, 2009_Westerkamp_PRL} and transition metal ferromagnets.\cite{2010_Sara_PRL,2011_Schroeder_JPCM}
All these  experimental observations of quantum Griffiths phases are in ferromagnets
rather than in antiferromagnets. However, in contrast to antiferromagnets, a complete theory of quantum Griffiths
phases in  ferromagnetic metals does not yet exist.

In this paper, we therefore
develop the theory of  quantum Griffiths effects  in ferromagnetic metals with both Ising and Heisenberg symmetries.
 We show that the  quantum Griffiths singularities do not take  power-law form, in contrast to those  in antiferromagnets.\cite{2006_Vojta_JPhysA, 2010_Vojta_JLTPhys}
The  rare-region density of states   behaves as  $\rho(\epsilon)\sim \exp[\{-\tilde{\lambda}\log (\epsilon_0/\epsilon)\}^{3/5}]/\epsilon$
in the low-energy limit,
where $\tilde{\lambda}$ plays a role analogous to the non-universal Griffiths exponent. This means that the Griffiths singularity is stronger
than a pure power law.  This kind of density of states leads to non-power-law dependencies
on the temperature $T$ of various observables, including the specific heat, $C\!\sim \exp[\{-\tilde{\lambda} \log(T_0/T)\}^{3/5}] $,
and the magnetic susceptibility, $\chi\!\sim \! \exp[\{-\tilde{\lambda} \log(T_0/T)\}^{3/5}]/T$.
The zero-temperature magnetization-field curve behaves as $M\sim \exp[\{-\tilde{\lambda} \log(H_0/H)\}^{3/5}]$.

The paper is organized as follows. In Sec. II, we introduce the model: Landau-Ginzburg-Wilson order
parameter field theories for ferromagnetic Ising   and Heisenberg  metals. In Sec. III,
we study the dynamics of a single rare region. For the Ising case, we use a variational instanton calculation, and for Heisenberg symmetry, we
use a renormalization
group theory of the quantum nonlinear sigma model with a damping term. In Sec. IV, we average over all rare regions and calculate observables in the ferromagnetic quantum Griffiths phase. In Sec. V, we compare our
predictions with existing experimental data.
Finally, we conclude in Sec. VI by discussing the difference between ferromagnetic and antiferromagnetic quantum Griffiths singularities
as well as some open questions.

\section{The model}

Rare region effects in disordered  metallic systems are realized both
in Ising magnets \cite{2000_Castro_PRB}  and in Heisenberg magnets. \cite{2005_Vojta_PRB} In the following, we consider both cases.
Our starting point is a quantum Landau-Ginzburg-Wilson  action of the itinerant  ferromagnet \cite{1976_Hertz_PRB,1993_Millis_PRB}$^,$
\footnote{We set Planck's constant and Boltzmann constant to unity ($\hbar=k_{B}=1$) in what follows.}
\begin{equation}\label{S}
S=S_{\rm{stat}} +S_{\rm{diss}} +S_{\rm{dyn}}\,,
\end{equation}
where the static part has the form
\begin{align}\label{SStat}
S_{\rm{stat}}=& E_0  \int_{0}^{\beta} d \tau \int d^{3}{\bf{r}} \Bigl[ t \phi^{2}({\bf{r}},\tau)
+  [\nabla \phi({\bf{r}},\tau)]^2\nonumber \\
&+ \frac{1}{2}
\phi^{4}({\bf{r}},\tau)  \Bigr]\,.
\end{align}
Here,  $E_0$ is a characteristic energy (assumed to be
of the order of the band width in a transition metal compound or the order of the Kondo-temperature
in an $f$-electron system). We measure lengths in units of the microscopic length scale $\xi_0$.
$t>0$ is the bare distance of the bulk system from  criticality.
$\phi({\bf{r}},\tau)$ is the dimensionless  order parameter field.
It  is a scalar for the Ising model,
while it has three components $(\phi_1, \phi_2, \phi_3)$ for a Heisenberg magnet.

We consider  disorder coupled to the square of the order parameter. The corresponding action has the form
\begin{align}\label{diss}
S_{\rm{diss}}=&E_0  \int_{0}^{\beta} d \tau \int d^{3}{\bf{r}}\ V({\bf{r}}) \phi^{2}({\bf{r}},\tau)\,,
\end{align}
where $V({\bf{r}})$ is the disorder potential.

The dynamical part of Eq. (\ref{S}) is  $S_{\rm{dyn}}=S^{(1)}_{\rm{dyn}}+S^{(2)}_{\rm{dyn}}$,
where
\begin{equation}\label{SD1}
S^{(1)}_{\rm{dyn}}=E_0 \tau_m^2 \int_{0}^{\beta} d \tau \int d^{3}{\bf{r}}[\partial_{\tau} \phi({\bf{r}},\tau)]^2\,,
\end{equation}
corresponds to the undamped  dynamics of the system with the clean dynamical exponent $z=1$, while
\begin{equation}\label{SD2}
S^{(2)}_{\rm{dyn}}=\frac{\gamma T}{E_0}  \sum_{\omega_n}|\omega_n|\int d^3 \textbf{q}\
\frac{|\tilde{\phi}(\textbf{q},\omega_n)|^2}{|\textbf{q}|^a}\,,
\end{equation}
describes overdamped dynamics with conserved order parameter (clean dynamical exponent $z=2+a$),
which stems from the coupling to the conduction electrons.
In Eq. (\ref{SD1}), $\tau_m$ is a microscopic time, and in Eq.  (\ref{SD2}),  $\gamma$ parametrizes the
strength of the dissipation. $\tilde{\phi}(\textbf{q},\omega_n)$  is the Fourier transform of the order parameter $\phi({\bf{r}},\tau)$  in momentum and Matsubara frequency. The value of $a$ depends on the character of the electron motion in the system
and equals  1 or 2 for  ballistic and diffusive ferromagnets, respectively.

\section{Dynamics of a single rare region}

In this section, we study the dynamics of a single droplet  formed on a rare region of linear size $L$.
This means, we consider a single spherical defect of radius $L$ at the origin
with potential $V(r)=-V$ for $r<L$, and $V(r)=0$ otherwise. We are interested in the case
$V>0$, i.e., in defects that favor the ordered phase.

The effective dimensionality
of the model defined by Eq. (\ref{S}) is $d_{\rm{eff}}=3+z.$ Thus, the clean model (\ref{S}) is
above its upper critical dimension ($d_c=4$), implying that mean-field theory is  valid.
The  mean-field equation for a static order parameter configuration $\phi_0(r)$ is \cite{2001_Millis_PRL}
\begin{equation}
\nabla^2\phi_0(r)+[t+V(r)]\phi_0(r)+\phi^3_0(r)=0\,,
\end{equation}
with  solution
\begin{equation}\label{O1}
\phi_0(r) =
\left\{
\begin{array}{lr}
\hspace{0.2cm} \phi_0 \hspace{1.34cm} \text{for} \hspace{0.5cm} r<L \\
\frac{\phi_0 L}{r}e^{-rt^{1/2}}  \hspace{0.3cm}  \text{for} \hspace{0.5cm} r>L.
\end{array}
\right.
\end{equation}
This implies that the order parameter is approximately constant in the  region $r<L$
and decays outside of it.

To study the dynamics of the droplet, we start from the
variational instanton approach.\cite{2002_Millis_PRB} In the simplest case, the droplet maintains its
shape while collapsing and reforming.  In order to estimate the action  associated with
this process, we make the ansatz
\begin{equation}\label{An1}
\phi (r,\tau)=\phi'_0(r) \eta(\tau)\,.
\end{equation}
Here, $\phi'_0(r)$ must be chosen such that $\int d^3 {\bf{r}} \phi (r,\tau)$ is
time independent because of order parameter conservation in an itinerant ferromagnet.
This can be done by introducing $\phi'_0(r)=\phi_0(r)(1-Ar)$ such that the $\textbf{q}=0$
Fourier component is canceled.
$A$ is a constant to be determined. In the limit of a large rare region, $Lt >> 1,$ we find
\begin{equation}
\phi'_0(r)=\phi_0(r)\left(1-\frac{4}{3}\frac{r}{L}\right) \,.
\end{equation}
In the following subsections, using ansatz Eq. (\ref{An1}), we separately discuss the dynamics of the droplet
in itinerant Ising and Heisenberg ferromagnets.

\subsection{Itinerant Ising model}

We now  calculate  the tunneling rate between the ``up" and  ``down" states of a single rare region in an itinerant Ising ferromagnet
by carrying  out variational instanton calculations.\cite{2002_Millis_PRB, 2007_Hoyos_PRB}
To estimate the instanton action, we use the ansatz Eq. (\ref{An1}) (which
provides a variational upper bound for the instanton action)  with $\eta(\tau)= \pm 1$ for $\tau \rightarrow \pm \infty$. Inserting this ansatz into the action Eq. (\ref{S}) and integrating over the spatial variables  yields, up to constant prefactors,
\begin{equation}\label{SB}
S_{\rm{stat}}\sim L^3\int d\tau [-2\eta^2(\tau)+\eta^4(\tau)]\,,
\end{equation}
and
\begin{align}\label{SK}
S^{(1)}_{\rm{dyn}}\sim L^3\int d \tau
[\partial_{\tau} \eta(\tau) ]^2\,.
\end{align}

The part of the action corresponding to the overdamped dynamics becomes
\begin{equation}\label{SD}
S^{(2)}_{\rm{dyn}}=\frac{\alpha}{4} \int d\tau d\tau' \frac{d\eta}{d\tau}\frac{d\eta}{d\tau'}\log\frac{(\tau
-\tau')^2+\tau^2_m}{\tau^2_m} \,,
\end{equation}
where the dimensionless dissipation strength $\alpha\sim \gamma L^{3+a}$.
In order to estimate the action Eqs. (\ref{SB}) to (\ref{SD}), we make the variational ansatz
\begin{equation}\label{Ansatz}
\frac{d\eta}{d\tau}=\frac{2\theta(
\tau^2_0-4\tau^2)}{\tau_0} \,.
\end{equation}
Summing all contributions, we obtain the instanton action
\begin{equation}
S\sim  L^3/\tau_0+  L^3\tau_0+\gamma L^{3+a}\log(\tau_0/\tau_m)\,.
\end{equation}

Minimizing this action over the instanton duration gives $\tau_0\sim L^{-a}/\gamma$.
Correspondingly, the action is $S \sim \gamma L^{3+a}$.
Then, the bare  tunneling rate or tunnel splitting behaves as
\begin{equation}\label{Itunrate}
\epsilon\sim \exp(-S)\sim \exp({-{\rm{const.}}\times\gamma L^{3+a}})\,.
\end{equation}
Thus,  the  bare tunneling rate  decays exponentially with $L^{3+a}$ in the itinerant  Ising ferromagnet
unlike the tunneling rate in  the itinerant  Ising antiferromagnet,\cite{2001_Millis_PRL,2002_Millis_PRB}
which decays exponentially with  ${L^3}$. The extra factor $L^{a}$ can be understood as follows.
To invert the magnetization of a rare region of linear size $L$,  magnetization must be transported
over a distance of the order of $L$, because the order parameter conservation prevents local spin flips.
The rare region dynamics thus  involves modes with  wave vectors of the order of
$q\sim 1/L$. Since  the part of the action corresponding to the overdamped dynamics Eq. (\ref{SD2}) is inversely
proportional  to momentum $q^a$, we obtain an extra factor $L^{a}$ in the action Eq. (\ref{SD}).

Within renormalization group methods,\cite{1987_Legget_RMP} the instanton-instanton interaction renormalizes the zero-temperature tunneling rate to
\begin{equation}\label{ItunrateR}
\epsilon_{\rm{ren}}\sim \epsilon^{1/{(1-\alpha )}}\,.
\end{equation}
This implies that at zero temperature, the smaller  rare regions with $\alpha < 1$ continue to tunnel with a strongly reduced rate, while the larger rare regions ($\alpha >1$) stop to tunnel and behave classically, leading to super-paramagnetic behavior.

\subsection{Itinerant Heisenberg Model}

A particularly interesting case are  itinerant Heisenberg ferromagnets because  quantum Griffiths
phases have been  observed  experimentally in these systems.\cite{2010_Sara_PRL, 2009_Westerkamp_PRL, 2011_Schroeder_JPCM} We now
study  the dynamics of a single rare region in an itinerant Heisenberg ferromagnet. We make the
ansatz
\begin{eqnarray}\label{AH}
\phi (r,\tau)=\phi'_0(r) \bf{n}(\tau)\,,
\end{eqnarray}
Here, $\bf{n}(\tau)$ is a three-component unit vector.
After substituting Eq. (\ref{AH}) into the action Eq. (\ref{S}) and  integrating over the spatial variables, we obtain
\begin{align}\label{S_01}
S\sim g \tau_m^2 \int d \tau [\partial_{\tau} {\bf{n}}(\tau) ]^2+\frac{\alpha}{4}\int d \tau d\tau' \frac{{\bf{n}}(\tau)\cdot{\bf{n}}(\tau')}{(\tau-\tau')^2+\tau^2_m}\,,
\end{align}
where the  dimensionless coupling constant $g\sim L^3 $ and $\alpha\sim\gamma L^{3+a}$ as before.
Because there is no barrier in a system with continuous order parameter symmetry, the static part of
the action is constant. Therefore, we cannot solve the problem within the variational instanton approach. Instead,
rotational fluctuations must be taken into account.

We calculate the characteristic relaxation time of the rare region by a renormalization group  analysis
of the action Eq. (\ref{S_01}). As shown in the Appendix, for weak damping $\alpha \ll g$, there
are two different regimes, where the behaviors of the  relaxation times are different. Particularly,
for  energies $\omega$  larger than some crossover energy $\omega_c\sim \alpha /g$, undamped dynamics is dominant,  and  the relaxation time of the rare region has the form
\begin{equation}\label{grelax0}
\xi^{{\tau}}_{g}\sim  L^3\,,
\end{equation}
which leads to a power-law dependence of the rare-region characteristic energy on $L$,
\begin{equation}\label{gtunrat}
\epsilon\sim  L^{-3}\,.
\end{equation}

For energies $\omega \ll \omega_c$, overdamped dynamics dominates the system properties, and  the relaxation time of the rare region behaves as
\begin{equation}\label{garelax0}
\xi^{{\tau}}_{\gamma}\sim \exp[{{\rm{const.}}\times \gamma  L^{3+a}}]\,.
\end{equation}
This results in a characteristic energy of
\begin{equation}\label{gatunrat}
\epsilon\sim \exp[{{-\rm{const.}}\times \gamma  L^{3+a}}]\,.
\end{equation}
Thus, the behavior of the characteristic energy in the itinerant Heisenberg magnet is the   analogous to that  of the
tunneling rate in
the Ising model discussed above.

We  can now roughly  estimate the size $L_c$ of the rare region corresponding to the crossover of the two  regimes. By comparing Eqs. (\ref{grelax0}) and (\ref{garelax0}), we find for small $\alpha$:
\begin{equation}\label{RSize}
L_c\sim  [ \log({\rm{const.}}/\gamma)/\gamma]^{1/(3+a)}\,.
\end{equation}
For small rare regions, $L<L_c$ , the undamped dynamics dominates systems properties and the characteristic energy is given by Eq. (\ref{gtunrat}),  while for $L>L_c$, the damping term is dominant  and  the characteristic energy is determined by Eq. (\ref{gatunrat}).

For  large damping $\alpha \gg g$, the  overdamped dynamics dominates the system properties for all energies $\omega$.
Correspondingly, the characteristic energy is given by Eq. (\ref{gatunrat}).

\section{Observables}

In the last section, we have seen that metallic Ising ferromagnets display modified Griffiths
behavior at higher energies [Eq. (\ref{Itunrate})], while at asymptomatically low energies,  the rare regions freeze and lead to a smeared phase transition [Eq. (\ref{ItunrateR})].
For Heisenberg ferromagnets, we have found conventional behavior at higher energies [Eq. (\ref{gtunrat})], and
modified Griffiths behavior at   low energies  [Eq. (\ref{gatunrat})]. Correspondingly, we expect modified Griffiths singularities in thermodynamic quantities at low energies for itinerant Heisenberg ferromagnets, while for metallic Ising ferromagnets they should occur at higher energies.

In this section, we use the single-rare-region results of Sec. III to study the  thermodynamics in these ferromagnetic quantum Griffiths phases.
To do so, we need to estimate the rare-region density of states. By basic combinatorics (see, e.g., Refs. \onlinecite{2006_Vojta_JPhysA, 2010_Vojta_JLTPhys}), the probability
for finding an impurity-free   rare region of volume $L^{3}$ is $\mathcal{P}\sim \exp(-b L^3)$ with $b$ being a constant
that depends on the disorder strength. Combining this and Eq. (\ref{gatunrat}) gives the density of states
(of the Heisenberg system) in the low-energy regime as
\begin{equation}\label{Density0}
\rho(\epsilon)\sim\frac{1}{\epsilon} \exp[\{ - \tilde{\lambda} \log(\epsilon_0/\epsilon)\}^{3/(a+3)}]\,.
\end{equation}
Here, $\epsilon_0$ is a microscopic energy scale, and the non-universal exponent  $\tilde{\lambda}\sim b^{(a+3)/3}/\gamma$ plays a role similar to the
usual quantum Griffiths exponent. The same density of states follows from Eq. (\ref{Itunrate}) for
the higher-energy regime of the Ising model. Thus, in ferromagnetic metals, the rare-region density of states does not take power-law form,
in contrast to the one in  antiferromagnets.

We can now find observables using the rare-region density of states Eq. (\ref{Density0}).
The number $n$ of free rare regions at temperature $T$ behaves as
\begin{align}\label{numb}
n(T)& \sim\int d\epsilon \rho(\epsilon) e^{-\epsilon /T}/(1+e^{-\epsilon /T}) \nonumber \\
& \sim \exp[\{-\tilde{\lambda} \log(T_0/T)\}^{3/(a+3)}] \,,
\end{align}
where $T_0$ is a microscopic temperature scale.

The uniform static susceptibility can be estimated by summing Curie susceptibilities for all free rare regions, yielding
\begin{eqnarray}\label{SusT}
 \chi(T)=n(T)/T\! \sim \! \frac{1}{T}\exp[\{-\tilde{\lambda} \log(T_0/T)\}^{3/(a+3)}]\,.
\end{eqnarray}
The dependence of the moment $\mu$ of the rare region on its energy leads to a subleading correction only.

The contribution of the rare regions to the specific heat $C$ can be obtained from
\begin{align}\label{Ene}
\Delta & E \sim\int d\epsilon \rho(\epsilon) \epsilon\ e^{-\epsilon /T}/(1+e^{-\epsilon /T})\nonumber \\
&\sim T \exp[\{-\tilde{\lambda} \log(T_0/T)\}^{3/(a+3)}] \,,
\end{align}
which gives $\Delta C\sim \exp[\{-\tilde{\lambda} \log(T_0/T)\}^{3/(a+3)}]$. Knowing the specific heat, we can find
the rare region contribution to the entropy  as $\Delta S\sim \exp[\{-\tilde{\lambda} \log(T_0/T)\}^{3/(a+3)}]$.

To determine the zero-temperature magnetization in a small ordering field $H$, we note that rare regions
with $\epsilon<H$ are (almost) fully polarized  while the rare regions with $\epsilon >H$ have very small
magnetization. Thus,
\begin{align}\label{mag}
m \sim\int_0^H d\epsilon \rho(\epsilon)\sim \exp[\{-\tilde{\lambda} \log(H_0/H)\}^{3/(a+3)}]\,,
\end{align}
where $H_0$ is a microscopic field
(again, the moment of  the rare region leads to a subleading correction).
The zero-temperature dynamical susceptibility  can be obtained by summing the susceptibilities of the individual
rare regions using the density of states Eq. (\ref{Density0}),
\begin{eqnarray}\label{Sus1}
 \chi(\omega)=\int_{0}^{\Lambda}d\epsilon \rho(\epsilon)  \chi_{\rm{rr}}(\omega; \epsilon)\,,
\end{eqnarray}
 where the dynamical susceptibility of a single  rare region in  Heisenberg metals  at zero temperature is given by \cite{2009_Vojta_PRB}
\begin{eqnarray}\label{Sus2}
\chi_{\rm{rr}} (\omega+i0; \epsilon)=\frac{\mu^2 }{\epsilon
- i\gamma\omega}\,,
\end{eqnarray}
where $\mu$  is the moment of the rare region. Substituting Eq. (\ref{Sus2}) into Eq. (\ref{Sus1}) we find
\begin{align}\label{Sus}
\chi (\omega\!+\!i0)\! \sim\! \frac{(1\!+\!i\gamma\ \! \rm{sgn}(\omega))}{|\omega|}\exp [\{-\tilde{\lambda} \log|\omega_0/\omega |\}^{3/(a+3)}]\,,
\end{align}
where $\omega_0$ is a microscopic frequency.
This result can be used to estimate the rare region contribution to the NMR spin relaxation time $T_1$.
Inserting Eq. (\ref{Sus}) into Moriya's formula \cite{1963_Moryia_JPSJ} for the relaxation rate  yields
\begin{eqnarray}\label{Rate}
1/T_1\sim \frac{T}{\omega^2} \exp[\{-\tilde{\lambda} \log|\omega_0/\omega |\}^{3/(a+3)}]\,.
\end{eqnarray}

\section{Experiment}
\begin{figure}[t]
\includegraphics[width=8.7cm,clip]{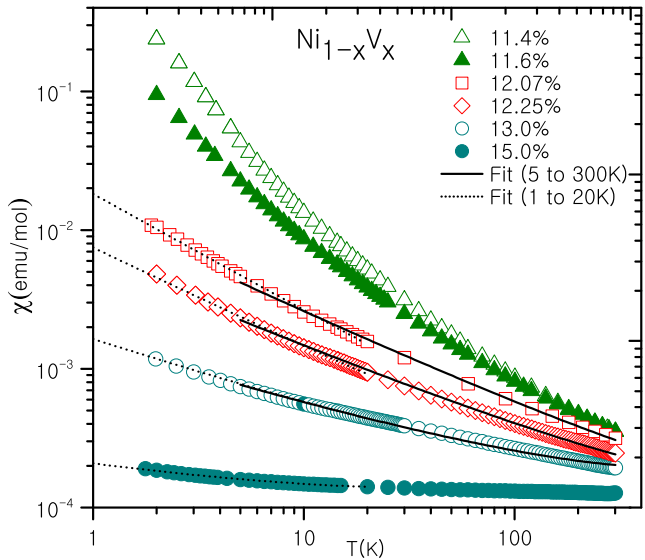}
\caption{(Color online). Temperature dependence of the susceptibility of ${\rm{Ni}}_{x}{\rm{V}}_{1-x}$ for  different Vanadium concentrations.
Solid  and dotted lines represent fits to  Eq. (\ref{SusT}) in the different temperature ranges 5 to 300 K and 1 to 20 K, respectively (data from Ref. \onlinecite{2010_Sara_PRL}).}
\label{fig:chi}
\end{figure}
Recently, indications of a   quantum Griffiths phase
have been observed in the transition metal ferromagnet ${\rm{Ni}}_{x}{\rm{V}}_{1-x}$. \cite{2010_Sara_PRL, 2011_Schroeder_JPCM}
The behavior of the  thermodynamics has been described  well in terms of the power-low quantum Griffiths singularities
predicted for an itinerant antiferromagnet (and the transverse-field Ising model).
Here, we compare our new theory of ferromagnetic quantum Griffiths phases with the experimental data given in Refs. \onlinecite{2010_Sara_PRL, 2011_Schroeder_JPCM}. The residual resistivity of  ${\rm{Ni}}_{x}{\rm{V}}_{1-x}$ close to the quantum phase transition is rather high.
\footnote{A. Schroeder, private communications} Thus, we choose $a=2$ for a diffusive ferromagnet.
Figure \ref{fig:chi} shows the  behavior of the susceptibility as a function of temperature.
The curves corresponding to the concentrations $x=13.0\%$ and $x=15.0\%$ (which are far away from the critical concentration $x_c\approx 11.5\%$) are described  better by power laws rather than our modified quantum Griffiths behavior Eq. (\ref{SusT}), at least above $T\approx 10\rm{K}$ (the low-temperature upturn is likely due to freezing of the rare regions).  For concentrations  $x=12.07\%$
and $x=12.25\%$, our theory fits better than power-law Griffiths singularities and extends the fit range from 30--300K down to 5--300K.
The  curves corresponding to the concentrations  $x=11.4\%$
and $x=11.6\%$ can be fitted by Griffiths power-laws only in the temperature range 30 to 300 K,
our new functional  form Eq. (\ref{SusT}) does not improve the fit of these curves.

We also compared the prediction Eq. (\ref{mag}) for  a modified magnetization-field curve with the data given in Refs.        \onlinecite{2010_Sara_PRL, 2011_Schroeder_JPCM}. We found that the  fits
to  power-laws and to the modified quantum Griffiths behavior Eq. (\ref{mag}) cannot be distinguished.

Let us also point out that the susceptibility data in the temperature range below 20K can be fitted
reasonable well by Eq. (\ref{SusT}); see details in Fig. \ref{fig:chi}.
Further experiments may be necessary to decide whether our theory applies in this region.

Overall, our theory does not significantly improve the description of the data of Refs. \onlinecite{2010_Sara_PRL, 2011_Schroeder_JPCM} over the
temperature range where Griffiths behavior is observed.
A possible reason is  that the relevant rare regions  are too small. At concentrations $x=13.0\%$ and $x=15.0\%$, they have moments of about  $\mu\approx 5\mu_{\rm{B}}$  and  $\mu\approx 1\mu_{\rm{B}}$, respectively. Correspondingly, the effect of the order parameter transport cannot play any role,
whereas our functional forms arise for large rare regions where the order parameter transport limits the relaxation of the rare region.
A possible reason why the curves corresponding to the concentrations  $x=11.4\%$
and $x=11.6\%$ can not be  described by our theory at $T<30\rm{K}$ might be
that the curves are actually slightly on the ordered side of the quantum phase transition.

\section{Conclusions}

In summary, we  studied the dynamics of  rare regions in disordered   metals
close to the ferromagnetic quantum phase transition, considering the cases of both Ising and Heisenberg spin symmetries.
The overall phenomenology is similar to the well-studied antiferromagnetic quantum Griffiths behavior. \cite{2000_Castro_PRB,
2002_Millis_PRB, 2003_Vojta_PRL, 2005_Vojta_PRB} Namely, for Ising symmetry
 at low temperatures, the overdamping causes sufficiently large
rare regions to stop tunneling. Instead, they behave classically, leading to super-paramagnetic behavior
and a smeared quantum phase transition. In contrast,
at higher temperatures but below a microscopic cutoff scale, the damping is unimportant and
quantum Griffiths singularities can  be observed.
In contrast to the Ising case, the
itinerant Heisenberg ferromagnet displays  quantum Griffiths singularities when damping is sufficiently strong, i.e., at low temperatures.
Above a crossover temperature,  conventional behavior is expected.

Although the phenomenologies of the ferro- and antiferromganetic cases are similar, the functional forms of the
quantum Griffiths singularities are different. In ferromagnetic quantum Griffiths phases, the tunneling rate
(or characteristic energy) of a rare region   decays
as $\exp[-{\rm{const.}}\times\gamma L^{a+3}]$ with  its linear size $L$,  where $a$ is equal to 1 or 2
for  ballistic and diffusive ferromagnets, respectively. This leads to the modified  nonpower-law quantum Griffiths singularities in thermodynamic quantities, discussed in Sec. IV, in contrast to the power-law quantum Griffiths singularities in itinerant antiferromagnets.
The reason is the following.  Because of the order parameter conservation
in the itinerant quantum ferromagnet, the damping effects are further enhanced as the
dimensionless dissipation strength $\alpha$ for a rare region of linear size $L$ is proportional
to $L^{a+3}$ rather than $L^3$.

In strongly disordered system, where our theory is most likely to apply, the motion of the electron is diffusive. Correspondingly, we expect $a=2$. In hypothetical
systems with rare regions, but   ballistic dynamics of the electrons, $a$ would take the value 1.

In our explicit calculations, we have used Hertz's form \cite{1976_Hertz_PRB}
of the order-parameter field theory of the itinerant ferromagnetic quantum
phase transition. However, mode-coupling effects in the Fermi liquid lead
to an effective long-range spatial interaction between the order parameter
fluctuations.
\cite{1996_PRB_Kirkpatrick, 1996_Vojta_EPL, 1997_Vojta_ZPB}
In the order-parameter field theory, this leads to a nonanalytic momentum dependence
of the static action Eq. (\ref{SStat}).  The effects of this long-range interaction
on the existence and energetics of a locally ordered rare region were studied in detail
in Ref. \onlinecite{2007_Hoyos_PRB}.
This work showed that the long-range interactions only produce subleading
corrections to the droplet-free energy. Therefore, including these
long-range interactions in the action Eq. (\ref{S}) will not change the results of the
present paper.

Let us now turn to the limitations of our theory.
In our calculations, we assumed that  the droplet
maintains its shape while collapsing and reforming. Correspondingly, our calculation provides a variational
upper bound for the instanton action. There could be faster relaxation processes;
however, it is hard to image the droplet dynamics  to avoid the restriction
coming from the order parameter conservation.
We treated the individual, locally ordered rare regions
as independent. But, in a real metallic magnet, they are weakly coupled by a  Ruderman-Kittel-Kasuya-Yosida (RKKY), interaction
which is not included in the Landau-Ginzburg-Wilson action Eq. (\ref{S}). At the lowest temperatures, this RKKY interactions between the rare regions
induces a cluster glass phase. \cite{2005_Miranda_PRL} Finally, our theory does not take the
feedback of the order parameter fluctuations on the fermions into account. It has been found that for some quantum phase transitions, the Landau-Ginzburg-Wilson theory breaks down sufficiently close to the transition point due to this feedback. \cite{2003_Abanov_AP, 2005_Belitz_RMP}
For strongly disordered systems, this question has not been addressed yet, it remains a task for the future.

Turning to experiment, our theory does not significantly  improve the description of the data of
${\rm{Ni}}_{x}{\rm{V}}_{1-x}$. \cite{2010_Sara_PRL, 2011_Schroeder_JPCM}
We believe that the main reason is that our theory is valid for asymptomatically
large rare regions where the order parameter transport plays an important role, whereas the experimental accessible rare regions in
${\rm{Ni}}_{x}{\rm{V}}_{1-x}$ are not large enough for the order parameter
conservation to dominate their dynamics. We expect our theory can be applied in systems
where one can observe Griffiths singularities at lower temperatures leading to larger rare regions.

\appendix*

\section{Renormalization group theory}

In this Appendix, we show the derivation of Eqs. (\ref{grelax0}) and (\ref{garelax0}) by  renormalization group (RG) analysis.
At low temperatures, the action Eq. (\ref{S_01})  is formally equivalent to a quantum non-linear sigma model
\cite{1977_Nelson_PRB} in imaginary time $\tau$.
We can set $\bf{n}(\tau)=(\pi(\tau),\sigma(\tau))$, where ${\pi}(\tau)=(\pi_1(\tau),\pi_2(\tau))$ represents transverse
fluctuations. After expanding in $\pi$
and keeping terms up to $\mathcal{O}(g^{-2})$, $\mathcal{O}(\alpha^{-2})$, we find \cite{1977_Nelson_PRB}
\begin{widetext}
\begin{align}\label{S_00}
S=\int \frac{d\omega}{2\pi} \ \left(g\omega^2 +\frac{\alpha}{4}|\omega|\right)
|\tilde{\pi}(\omega)|^2 +
\int \frac{d\omega_1 d \omega_2 d \omega_3 }{(2\pi)^3}\left(\frac{\alpha}{8}|\omega_1|-g\omega_1\omega_3\right) \tilde{\pi}_{\beta}(\omega_1)\tilde{\pi}_{\beta}(\omega_2)\tilde{\pi}_{\beta'}(\omega_3)
\tilde{\pi}_{\beta'}(-\omega_1-\omega_2-\omega_3) \,.
\end{align}
\end{widetext}

We now consider the case of the small damping $\alpha \ll g$. Two different energy regimes  can be distinguished: (i)  $\omega$ larger
than some crossover energy $\omega_c\sim\alpha /g$, implying that the undamped dynamics dominates the
systems properties,  and (ii)   $\omega\ll \omega_c$, when the damping term is dominant.

(i) Because  the contribution of the undamped dynamics is dominant in this regime, we neglect the damping term and  renormalize $g$.
 To construct a perturbative renormalizaition group, consider a  frequency region  $[-\Lambda,\Lambda]$ ($\Lambda$ is a high energy cut off),
and divide the modes into slow and fast ones, $\tilde{\pi}(\omega)=\tilde{\pi}^{<}(\omega)+\tilde{\pi}^{>}(\omega)$. The modes $\tilde{\pi}^{<}(\omega)$
involve frequency $-\Lambda/b<\omega<\Lambda/b$, and  are kept. We integrate out the short-wavelength
fluctuations $\pi^{>}(\omega)$ (with frequencies in the region $-\Lambda<\omega<-\Lambda/b$ and
$\Lambda/b<\omega<\Lambda$)  in perturbation theory using the propagator $
{\langle\tilde{\pi}^{>}_{\beta}(\omega)\tilde{\pi}^{>}_{\beta'}(\omega')\rangle} =
{\pi\delta_{\beta \beta'}\delta(\omega+\omega')}/({g \omega^2})$.

After applying standard techniques, we find that this coarse graining changes
the coupling constant $g$ to $g_{\rm{co}}=g+I_{g}(b)$,
where $I_{g}(b)=(2\pi\Lambda)^{-1}(b-1)$. After rescaling $\tau'=\tau/b$ and renormalizing
$\pi'(\tau')=\pi^{<}(\tau)/\zeta_{g}$, we obtain the renormalized coupling constant in the form
\begin{equation}\label{GR}
g'=b^{-1}\zeta^2_{g}g_{\rm{co}}\,.
\end{equation}
To find the rescaling factor $\zeta_{g}$, we average ${\mathbf{n}}$ over the short wavelength
modes $\pi^{>}$ and obtain
\begin{align}\label{nG}
\langle {\bf{n}} \rangle^>=&\langle(\pi^{<}_1+\pi^{>}_1,...,\sqrt{1-(\pi^{<}+\pi^{>})^2})\rangle^> \nonumber \\
=&(1-\langle (\pi^{>})^2\rangle^>/2 +\mathcal{O}(g^{-2}))(\pi^{<}_1,...,\sqrt{1-(\pi^{<})^2})\,.
\end{align}
Thus, we identify
\begin{align}\label{Zg}
\zeta_{g}=1-\langle (\pi^{>})^2\rangle^>/2 +\mathcal{O}(g^{-2})= 1-\frac{I_{g}(b)}{g}+\mathcal{O}(g^{-2})\,.
\end{align}
Correspondingly, the renormalized coupling constant given in Eq. (\ref{GR}) becomes
\begin{equation}\label{GR1}
g'= b^{-1}(g-I_{g}(b))\,.
\end{equation}
Setting  $b=1+\delta l$, and integrating  Eq. (\ref{GR1}) gives  the recursion relation $g(l)=g(0)e^{-l}$.
To find the  relaxation time, we run the RG to $g(l)=1$ and use $\xi^{{\tau}}\sim e^l$. This gives
\begin{equation}\label{grelax}
\xi^{{\tau}}_{g}\sim L^3\,.
\end{equation}

(ii) In the same way, for low  energies $\omega \ll\omega_c$,  we neglect the term
corresponding to the undamped dynamics and  renormalize the $\alpha$ coefficient.   We find that $\alpha$ is not modified
by the perturbation, i.e., $\alpha_{\rm{co}}=\alpha$, and the field rescaling factor $\zeta_{\alpha}$
is given by
\begin{align}\label{Zgamma}
\zeta_{\alpha}=1-\frac{I_{\alpha}(b)}{\alpha}+\mathcal{O}(\alpha^{-2})\,,
\end{align}
where $I_{\alpha}(b)=2\pi^{-1}\log(b)$. Then, we find the recursion relation ${\alpha}(l)={\alpha}(0)-4\pi^{-1}l$.
This leads to  the relaxation time
\begin{equation}\label{garelax}
\xi^{{\tau}}_{\gamma}\sim  \exp[{{\rm{const.}}\times\gamma  L^{3+a}}]\,.
\end{equation}

\section*{Acknowledgements}
This work has been supported by the NSF under Grant No. DMR-0906566.

\bibliographystyle{apsrev4-1}
\bibliography{C:/Users/dn9z2/Documents/Research/Res/texfile/b}

\end{document}